\documentclass[fleqn]{elsart}
\usepackage{amsmath,graphicx}
\begin{document}

\begin{frontmatter}

\title{Time series analysis for minority game simulations of financial markets}

\author{Fernando F. Ferreira},
\author{Gerson Francisco\corauthref{gerson}},
\ead{gerson@ift.unesp.br}
\author{Birajara S. Machado}, and
\author{Paulsamy Muruganandam}
\ead{anand@ift.unesp.br}
\address{Instituto de F\'{\i}sica Te\'orica, Universidade Estadual
Paulista, 01405-900 S\~ao Paulo, S\~ao Paulo, Brazil}
\corauth[gerson]{Corresponding author}
\date{\today}

\begin{abstract}

The minority game (MG) model introduced recently provides promising insights into
the understanding of the evolution of prices, indices and rates in the 
financial markets. In this paper we perform a time series analysis of the model
employing tools from statistics, dynamical systems theory and stochastic
processes. Using benchmark systems and a financial index for comparison, several  
conclusions are obtained about the generating mechanism for this kind of  evolution.
The motion is deterministic, driven by occasional random external perturbation. 
When the interval between two successive perturbations is sufficiently large, 
one can find low dimensional chaos in this regime. However, the full motion of 
the MG model is found to be similar to that of the first
differences of the SP500 index: stochastic, nonlinear and  (unit root)
stationary.

\end{abstract}
\end{frontmatter}

\section{Introduction}

The pricing of contingent claims contracts in financial economics is often
based on very  restrictive assumptions about the time evolution of the
underlying instrument \cite{black-scholes:73-01}. In recent years researchers
have endeavored to remove some of these restrictions by proposing more
realistic models which would incorporate features found in real markets
\cite{johnson:00-01,giardina:01-01,challet:01-01,lux:01-01}. There is
however a trade-off between analytic tractability and adherence to stylized
facts observed from empirical financial time series. The most attractive
features of the usual Black-Scholes type of models is the  possibility of
obtaining closed, exact formulas for the premium of derivative securities, and
to build a risk free replicating strategy. Such qualities are amply used in
financial institutions which require fast calculation and tools to hedge risky
assets. 

However, the geometric Brownian motion assumption of the Black-Scholes models 
ignores several empirically observed features of the real markets such as,
volatility clusters, fat tails, scaling, occurrence of crashes, etc. It is as
yet unknown which stochastic process is responsible for the motion of risky
assets, but physicists have taken some important steps in the right 
direction\cite{stanley:book:00,jefferies:00-01,bouchaud:book:00,challet:01-01,farmer:97-01}. In this work we implement a microscopic
agents-based model of market dynamics which gives rise to a quite complex and
rich behaviour\cite{challet:98-01} and whose output are macroscopic quantities
such as price returns. By varying its parameters the model exhibits market
crashes, Gaussian statistics and short ranged correlation, fat tailed returns
and long range correlation. This model retains the nontrivial opinion formation
structure of the grand canonical minority game\cite{arthur:94-01,zhang:98-01}
because it incorporates two new features. The first one is to allow two
categories of agents, producers (who use the market for exchanging goods), and
speculators (whose aim is to profit from price fluctuations). The second
feature is that speculators might choose not to trade, and in this sense the
model is similar to the grand canonical ensemble  of statistical physics
since the number of active traders is not constant. 

We perform an analysis of the time series generated by this model in order  to
classify its dynamical behaviour. We first test the data for unit roots and
remove  a simple kind of nonstationarity by taking differences wherever
necessary. The BDS statistic \cite{brock:96-01}, originated in the chaos
literature, uses the correlation  integral \cite{grassberger:83-01} as the
basis of a test for the hypothesis that the  data is independent and
identically distributed ({\it i.i.d.}). We apply this statistic as a model 
specification test by applying it to the residue of an ARIMA, autoregressive  
integrated moving average, process (although  this might not remove all kinds
of linearity) \cite{box:book:94}. If the null hypothesis is rejected  then this
is indication that the data is nonlinear. Another test based on surrogate data
\cite{theiler:92-01} is used to confirm the nonlinearity of the model. Since
the alternative hypothesis for the BDS procedure is not specified other tests
have to be applied in order to determine whether the nonlinearity comes from a
stochastic or deterministic mechanism. Such distinction is subtle and here we 
approach this question by computing more  parameters, the correlation 
dimension and two other procedures which do not require embeddings: recurrence
plots \cite{eckmann:87-01,gilmore:93-01}, and the Lempel-Ziv complexity (LZC)
\cite{lempel:76-01,badii:book:99}. In complement to the above procedures we
implement a  Bayesian approach called cluster weighted modelling (CWM)
\cite{gershenfeld:99-01,gershenfeld:book:99} in order to find further
indication of  determinism. 

The paper is organized as follows. The market model and its trajectories are
analysed in Section \ref{sec:2}. The time series analysis, including all
statistical tests is discussed in  Section \ref{sec:3}. A succinct presentation
of the CWM and its results are found in Section \ref{sec:4}. Analysis  of the
results obtained and a classification of the model evolution are given in
Section \ref{sec:5} while in the last  section we summarize the main results
and comment on future work.  

\section{Market Model}\label{sec:2}

Inspired by the {\it El Farol Problem} proposed by Arthur \cite{arthur:94-01},
the so called Minority Game  (MG) model introduced by Challet and Zhang
\cite{challet:97-01,zhang:98-01} represents a fascinating  toy-model for
financial market. Now it is becoming a paradigm for complex adaptive systems,
in which individual members or traders repeatedly compete to be in the winning
group. The game consists of $N$ agents that  participate in the market buying
or selling some kind of asset, e.g. stocks.  At any given time agent $i$ can
take two possible actions $a_{i}=\pm 1$, meaning buy or sell.  Those players
whose bets fall in the minority group are the winners, i.e., the sellers win if
there is an excess of buyers, and  vice versa. To determine the minority group
we just consider the sign of the global action $A(t)=\sum_{i}a_{i}$,  so that
if $A(t)$ is positive the minority is the group of sellers; in this case the
majority of   players expect asset prices to go up. 

In other words, this dynamics follows the law of demand and supply. In the end
of each turn, the output is a single label, $1$ or $-1$, denoting the winning
group at  each time step. This output is made available to all traders, and it
is the only information they can use to make decisions in subsequent turns.
Indeed, they store the $m$ most recent output of winners set $\mu(t)$. In this
way a limited memory of length $m$ is assigned to the traders corresponding to
the most recent history bit-string that traders use to make decisions for the
next step. In order to decide what action to take, agents use strategies. A
strategy is an object that processes the outcomes of the winning sets in the
last $m$ bets and from this information it determines whether a given agent
should buy or sell for the next turn. When a tie results from the strategies, 
buying or selling is decided by coin tossing. The memory defines $D=2^{m}$ possible
past histories so the strategy of one agent can be viewed as a D-dimensional
vector, whose elements can be 1 or -1. The space $\Gamma$ of strategies is an
hypercube of dimension $D$, and the total number of strategies in this space is
$2^{D}$. At the beginning of the game each agent draws randomly a number $S$ of
strategies from the space $\Gamma$ and keeps them forever. After each turn, the
traders assigns one (virtual) point to each of the strategies which would have
predicted the correct outcome. Along the game the traders always choose the
strategy with  the highest score. The MG is a very simple model, capable of
exhibiting complex behaviour, like phase transitions between an
information-efficient phase and information-inefficient phase, by just varying
a control parameter $\alpha=\frac{2.2^{m}}{NS}$, e.g., the ratio between
information complexity and number of strategies present in the game.
\begin{figure}
\begin{center}
\includegraphics[width=0.6\linewidth]{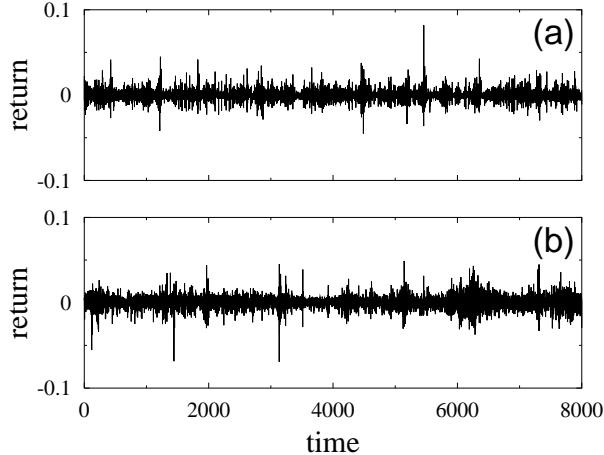}
\end{center}
\caption{Figure showing the returns for (a) Minority game model with
$N_p=1500$, $N_s=510$, $m=4$ and $\epsilon=0.003$ and (b) SP500 data}
\label{fig:return}
\end{figure}

More realistic features are reached with the grand canonical minority
game\cite{challet:01-01}. In this model we define the price process in terms of
excess demand $A(t)$,
\begin{equation} 
\log p(t+1)= \log p(t)+A(t),
\end{equation} 
and introduce two kinds of agents. 
\begin{itemize}

\item 

The first kind is called producers, who go to the market only for the 
purpose of exchanging goods; they only have one strategy and in this sense 
they behave in a deterministic  way with respect to $\mu (t)$. The number of
producers is $N_{p}$.

\item  

The other kind of agents are the speculators, who go to the market to make 
profit from price fluctuations. Since they are endowed with at least two
strategies during the game they need to use the best strategy; in this sense
speculators are adaptative agents with  bounded rationality.  The number of
speculators is $N_{s}$.

\end{itemize}   
In this version of the game, the number of speculators  can change anytime,
because the agent may decide not to trade in which case $a_{i}=0$. Since the
strategy chosen by speculators is the one with highest score, it is very
important to update the scores of each strategy at each time step. The
updating of the scoring $U_{i,s}$ of each strategy $s$ belonging to agent $i$
is given by the following equations
\begin{eqnarray}
U_{i,s}(t+1) & = & U_{i,s}(t) - a_{i,s}^{\mu(t)} A(t)\;\; \mbox{for}
\;\;s>0
\nonumber \\
U_{i,0}(t+1) & = & U_{i,0}(t) + \epsilon 
\end{eqnarray}
where $\epsilon $ is a threshold parameter. A sample of the trajectory
generated by this model is shown in Fig.~\ref{fig:return}(a). 
An important piece of information to understand the generating mechanism 
of the system can be obtained from the periods where no coin is tossed. 
The game is necessarily deterministic during this mode of evolution and 
we selected a rather long period of 1114 points to perform 
a time series analysis in this regime. 

This work is based upon the grand canonical MG model, henceforth called MG
model. 

\section{Time Series Analysis}\label{sec:3}

In this section we discuss the tools employed in time series analysis  starting
with the BDS statistics. We generate $10000$ points for each  of the benchmark
data described in the Appendix. Another benchmark is the set comprising $9998$
closing prices for the index SP500 from January 01, 1965 to January 01, 1995,
shown in Fig.~\ref{fig:return}(b). All time series used here are unit root
stationary with the exception of the financial index: this nonstationary is
removed by taking first difference.

The BDS test uses the correlation integral \cite{grassberger:83-01} as the
basis of a statistic  to test whether a series of data is {\it i.i.d.} In the
chaos literature the correlation integral is part of an efficient tool to
compute the fractal dimension of objects called attractors (for a formal
definition of attractor see, e.g., Ref.~\cite{milnor:85-01}). Given a sample of
empirical data $\{x_t\}_{t=1}^{N}$, the theory of  state-space reconstruction
\cite{takens:81-01} requires that the $d$-histories of the  data be
constructed, where $d$ is called  the embedding dimension. Under certain
conditions it is  possible to reproduce in this space the dynamics of the
system for a correct  choice of $d$. The correlation integral is a function
defined  on the trajectories in this space and from it one can compute the 
correlation dimension. A simple test for determinism consists of increasing 
the embedding dimension  and observing the occurrence of  a corresponding
increase in the correlation dimension.  Some conditions  have to be met in
order to apply this method, mainly stationarity and  sufficient number of data
points \cite{schreiber:97-01,gao:99-01,eckmann:92-01}. Our results show that 
the increase of the correlation dimension for the MG model with respect to the
embedding dimension is practically identical to the stochastic benchmark
series, including the index  SP500. In this work we resort to  better ways of
analysing the occurrence of stochastic behaviour in  complex time series
evolution.

The BDS statistic will be applied to the residue of ARIMA processes in order to
detect nonlinearity in the data. In this sense the statistic is used as a 
specification test. The asymptotic distribution of the statistic under the 
null of pure whiteness, is the standard normal distribution. The alternative 
hypothesis is not specified \cite{brock:96-01}. The code implemented here  is
taken from Ref.~\cite{lebaron:97-01}. From Table~\ref{table1}, the null
hypothesis for the MG model and for the SP500 index are  rejected more strongly
than for the nonlinear stochastic model $NLMA(2)$.    

In the surrogate data analysis, a null hypothesis is tested under a measure
$M$, usually some nonlinear statistic. Surrogates are copies of  the original
time series preserving all its linear stochastic structure. Let
$M_{\mbox{orig}}$ and $M_{\mbox{surr}}$ be the test measure for the original
and the surrogate data sets, respectively. The null hypothesis is rejected only
if $M_{\mbox{orig}} < \langle M_{\mbox{surr}} \rangle$ or $M_{\mbox{orig}} > 
\langle  M_{\mbox{surr}} \rangle$ for all surrogates. Some care should be taken
to test the null hypothesis, otherwise  false-positive rejection of null
hypothesis could result. A detailed analysis on this is found in
\cite{rapp:ijbc:2001:01}.      

Here surrogate data sets \cite{theiler:92-01,TISEAN} are used in complement to
the BDS statistic to test for nonlinearity. The null hypothesis is that the
data is described by a stationary linear stochastic process with Gaussian inputs. 
The test statistic $M$ is a simple measure of predictability as in 
\cite{kantz:book:97} and the procedure is to
generate $19$ samples of rescaled surrogates and to compare with the MG model.
If the test statistic falls outside the interval defined by the  surrogates,
then the null is rejected at the $95$\% level.  We found that, for embedding
dimensions from 2 to 5, the MG is nonlinear at this level of  significance,
corroborating the previous result using the BDS. 

In the remaining of this Section we discuss methods which do not require 
embeddings.
 
Another measure of randomness that provides further insight into time series
dynamics is the Lempel-Ziv complexity \cite{lempel:76-01,badii:book:99}.  No
embedding is necessary and the data is interpreted as a binary signal generated
by some kind of source. This idea is ever  present in communication theory
where one wishes to determine the minimum  alphabet required to code a source
whose signal is to be sent through a noisy channel. Let us consider the length
$L(N)$ of the minimal program  that reproduces a sequence with $N$ symbols. The
Lempel-Ziv algorithm is constructed by a special parsing which splits the
sequence into words of the shortest length and that has not appeared
previously. For example, the sequence $0011101001011011$ is parsed as
$0.01.1.10.100.101.1011$. One can  show that $L(N)\approx N_{w}(\log N_w(N)+1)$
where $N_w$ is the number of distinct words in a parsing and $N$ the size of
the sequence. From this one can see that $L(N)$ contains a measure of
randomness where a source that produces a greater number of new words is more
random than a source producing a more repetitive pattern. In analogy with
dynamical evolution, those  systems that are composed of well defined cycles
are predictable  while chaotic motion and stochastic processes are always
producing new kinds  of trajectories that never repeat themselves. A comparison
between chaos generated by differential equations and stochasticity  can now be
obtained. In the former case the Lempel-Ziv complexity is well  below 1 while
in the latter it is close to this value. More specifically, if we consider an
oscillatory system such as the well known van der Pol oscillator, then
$C=0.049$. For the Lorenz attractor with 2000 points $C=0.181$. In Section
\ref{sec:5} we comment on the discrepancy between  this value and that in
Table~\ref{table2}, and discuss the use of   this complexity measure in
dynamical systems generated by maps.   The complexity for the  MG dynamics  is
found to be $0.82$ while the financial index has $C\approx 1$.
\begin{figure}[!ht]
\begin{center}
\includegraphics[width=\linewidth]{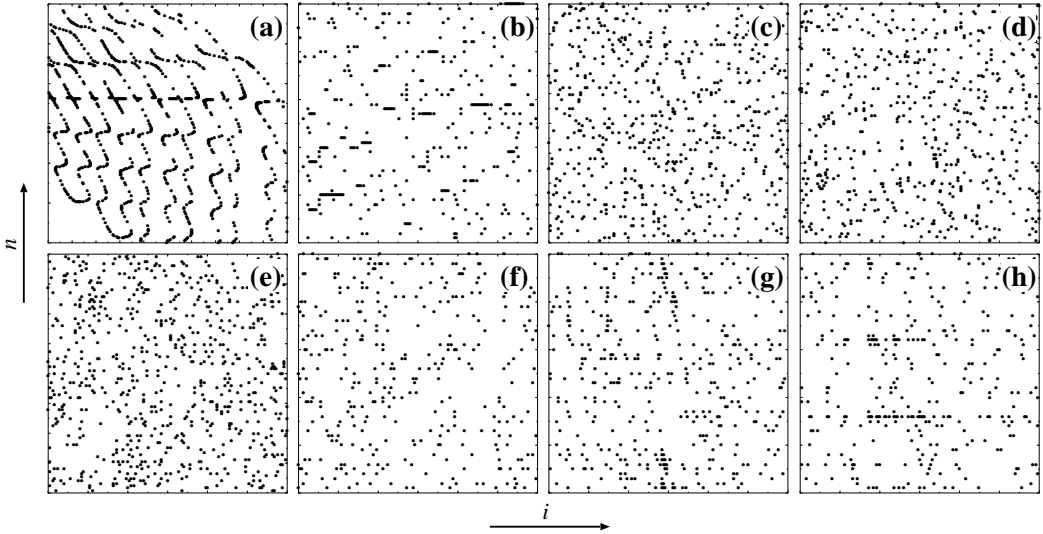}
\end{center}
\caption{Recurrence plot (a) Lorenz model, (b) Henon system, (c) random numbers
(d) ARMA(2,1), (e) NLMA(2), (f) SP500, (g) minority game model and (h) minority game
in the deterministic regime.}
\label{fig:rp}
\end{figure}

The implementation of recurrence plots \cite{eckmann:87-01}  used here is taken
from \cite{gilmore:93-01} since it provides  a clear distinction of the systems
we  intend to classify. The idea is to detect regions of ``close returns"  in a
data set. The construction of the plots is simple: just compute the absolute
values  $\mid x_{i}-x_{n}\mid$ for all the points in the data base. If the
horizontal axis is designated by $i$, corresponding to $x_{i}$, and the
vertical axis by $i+n$, corresponding to $x_{i+n}$, then  plot a black dot at
the site  $(i,n)$, whenever the absolute value difference is lower than
$\delta$; otherwise plot it white. Actually the time series is normalized to
$(0,1)$ and then $\delta$ is taken as $2-6$\% of the average distance between 
successive points. The black/white pattern can  be used to detect determinism
in the data. There is a clear difference  amongst plots generated by
differential equations, maps, random data and stochastic processes as  shown in
Fig.~ \ref{fig:rp}. Patterns  of horizontal segments in the recurrence plot
indicate the presence of  unstable periodic orbits in maps or differential
equations. In this sense, these  plots detect low dimensional chaos in
relatively small and even noisy data sets.

\section{Cluster-Weighted Modelling}\label{sec:4}

An interesting probability density estimation approach to characterize and
forecast time series  developed by Gershenfeld, Schoner and Metois
\cite{gershenfeld:99-01} is the so called cluster-weighted modelling. This
seems to be a powerful technique as it characterizes extremely well the time
series of nonlinear, nonstationary, non-Gaussian and discontinuous systems
using probabilistic dependence of local models. The cluster-weighted modelling
technique estimates the functional dependence of time series in terms of delay
coordinates. The main task of this approach is to find the conditional
forecast by estimating the joint probability density.
\begin{figure}[!ht]
\begin{center}
\includegraphics[width=\linewidth]{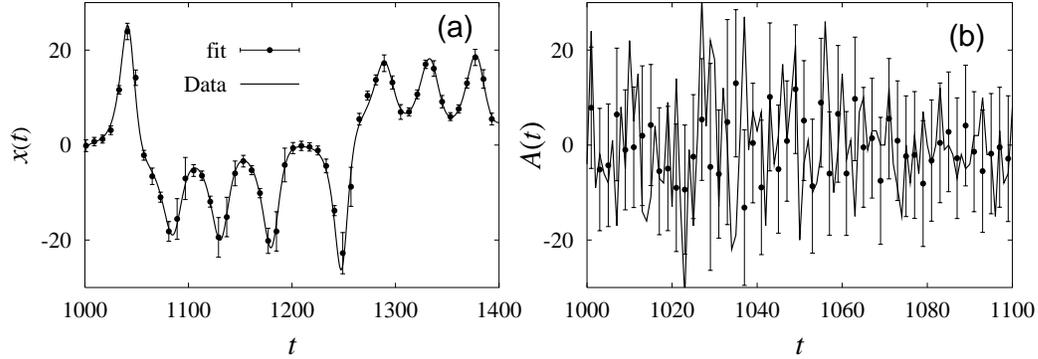}
\end{center}

\caption{Fitting of data using cluster weighted analysis (a) Lorenz system, $x$
component is fitted by $\langle y|\vec x \rangle$ where $y=x(t)$, $\vec x = 
\{x(t-\tau), x(t-2\tau), t\}$ with delay $\tau$ and (b) minority game model,
return is fitted in a similar fashion as in (a)}

\label{fig:cwm}
\end{figure}

Let $\{y_n,\vec x_n\}_{n=1}^N$ be the $N$ observations in which $\vec x_n$ are
known inputs and $y_n$ are the corresponding outputs. By knowing the joint
probability density $p(y,\vec x)$, we can derive the conditional forecast, the
expectation value of $y$ given $\vec x$, $\langle y|\vec x \rangle$. We can
also deduce other quantities such as the variance of the above estimation. 
Actually, the joint density $p(y,\vec x)$ is expanded terms of clusters which
describe the local models. Each cluster contains three terms namely, the weight
$p(c_m)$, the domain of influence in the input space $p(\vec x|c_m)$, and
finally the dependence in the output space $p(y|\vec x, c_m)$. Thus the joint
density can be written as \cite{gershenfeld:book:99},
\begin{align}\label{eqn:joint_den}
p(y,\vec x) = \sum_{m=1}^{M} p(y|\vec x, c_m)\, p(\vec x|c_m)\, p(c_m)
\end{align}
Once the joint density is known the other quantities can be derived from
$p(y,\vec x)$. For example, the conditional forecast is given by,
\begin{align}\label{eqn:expect_y}
\langle y|\vec x \rangle = \int y\, p(y|\vec x)\,dy 
= \frac{\sum_{m=1}^{M} f(\vec x, \beta_m)\,p(\vec x|c_m)\,p(c_m)}
{\sum_{m=1}^{M} p(\vec x|c_m)\,p(c_m)}.
\end{align}
Here $f(\vec x, \beta_m)$ describes the local relationship between $\vec x$ and
$y$. The parameters $\beta_m$ are found by maximizing the cluster-weighted
log-likelihood. The simplest approximation for the local model is with linear
coefficients of the form,
\begin{align}\label{eqn:linear_model}
f(\vec x, \beta_m) = \sum_{i=1}^{I}\beta_{m,i} f_i(\vec x).
\end{align}
The method just described is capable of modelling a wide range of 
deterministic time series. Here we use cluster weighted modelling to 
distinguish between deterministic and stochastic time series. In deterministic
systems we  observe that the variances of the different clusters converge to
values lower  than the variance  of  the original time series and one can
verify that  this property is robust under  changes of the number of clusters.
On the other hand stochastic systems do not have this property.
Fig.~\ref{fig:cwm} illustrates the comparison between the fitting of a
deterministic system (Lorenz) and the minority game data using cluster weighted
modelling.

\section{Analysis and Classification}\label{sec:5}

The main objective is to understand the  minority game mode of evolution and
other similar time series behaviour. Although this  system is not generated  by
any kind of differentiable dynamics or even  stochastic differential
equations, we use in its analysis  methods from dynamical systems, stochastic
processes and complex  systems theory. Tables~\ref{table1} \& \ref{table2} and
Fig.~\ref{fig:rp} summarize the main results.  We will make frequent reference
to the SP500 index  since the minority game model is supposed to reproduce the
dynamical  evolution of financial markets and this index is used as a
benchmark  for comparison. 

\begin{table}[!ht]

\caption{BDS statistic}

\label{table1}
\begin{minipage}{\linewidth}
\begin{tabular}{llll}
\hline
Data set & Epsilon & BDS-statistic  & Decision \\ 
         & ($\varepsilon$)\footnote {The value of $\varepsilon$ 
           is taken as one-half the deviation of the data set.} 
	           & (Embedding dimension: 2-8) & \\ 
\hline
Lorenz system & 0.070 & 610.2067, 1031.9777, 1955.1017  & strongly reject \\
& &  4071.2547, 9083.3612, & linearity  \\
& &  21286.7083, 51751.7514 & \\ 
Henon map & 0.300 & 101.1437, 183.6664, 368.5771, & strongly reject \\ 
& & 716.3989, 1474.7676,& linearity\\ 
& & 3129.6945, 6832.0170 & \\
Random & 0.500 & -0.5487, -0.4410, -0.3585, & accept linearity\\
& & -0.6511, -0.8622, & \\ 
& & -0.2150, 0.3504 & \\
ARMA(2,1) & 0.500 & -0.0270, -0.5267, -0.6978,  & accept linearity\\
& & -1.0838, -2.0321, & \\ 
& & -2.6126, -2.5441 & \\
NLMA(2) & 0.600 & 6.1692, 11.6768, 13.3927, & reject  linearity\\
& & 14.3183, 14.5538, & \\ 
& & 14.8573 15.6130 & \\
SP500 & 0.003 & 15.4909, 19.8118, 23.6680, & reject  linearity\\
& & 27.3441, 33.1618, & \\ 
& & 40.8782, 50.6347 & \\
Minority Game & 2.800 & 14.3499, 18.2687, 22.3293, & reject  linearity\\
& & 26.6340, 31.2736, & \\ 
& & 36.8712, 43.6115 & \\
Minority Game\footnote{Data set with length 1114 between two successive coin tossings.}
& 0.03 & 4.8767, 5.1903, 7.3879 &  reject  linearity\\
& & 12.3027, 19.1270, & \\ 
& & 31.8882, 60.1133 & \\
\hline
\end{tabular}
\end{minipage}
\end{table}

The BDS statistic and surrogate data analysis provide clear evidence that the
minority game is nonlinear.  The benchmark systems described  in the
Appendix reproduce the expected results for the BDS in Table~\ref{table2} and
we can clearly see that the null hypothesis is strongly rejected for known
deterministic nonlinear systems. As for the nonlinear stochastic system, the
index and the MG model, the rejection occurs at all dimensions and the
probability of type I error is practically zero.  In particular, the BDS was
used in other instances of the MG model, using different parameters and time
intervals with the same result. These findings  support the conclusion that
there is a nonlinear mechanism in operation which drives the MG dynamics and
that this property is robust to the extent tested herein. 
\begin{table}[!ht]
\caption{Summary of test results}
\label{table2}
\begin{minipage}{\linewidth}
\begin{tabular}{llllll}
\hline
Data set & Unit & ARIMA   & BDS\footnote{See details in Table~\ref{table1}.}& LZC & Recurrence \\ 
         & Root & (i,j,k) &  & $(C)$ & plot \\ 
\hline
Lorenz system & stationary & $2, 0, 0$ & nonlinear & 0.0677 & chaotic  \\ 
Henon map &  stationary & $4, 0, 3$  & nonlinear & $0.5754$ & chaotic \\
Random & stationary  & $0, 0, 0$  & linear & $\approx 1$ &  random\\
ARMA(2,1) & stationary & $2, 0, 0$  & linear & $0.8451$ & random \\
NLMA(2) & stationary & $0, 0, 2$  & nonlinear & $\approx 1$ & random \\
SP500 & non-stationary & $1, 0,1$ & nonlinear & $\approx  1$ & random \\
Minority Game & stationary & $1, 0, 5$ & nonlinear & $0.82$ & random \\
Minority Game\footnote{Data set with length 1114 between two successive coin tossings.}
              & stationary & $1, 0, 0$ & nonlinear & $0.6$ & chaotic \\
\hline
\end{tabular}
\end{minipage}
\end{table}

The Lempel-Ziv complexity is an important parameter that can be used in the
analysis of complex systems. Its advantage is that it does  not require
embeddings and can be easily employed in conjunction with other methods. The
results in Table~\ref{table1} supports the idea that there is a 
stochastic mechanism in
operation driving the MG model. There is a higher degree of indeterminacy in
the SP500 and this is  perhaps due to the fact that in this index there is a
certain amount of measurement noise. The surprisingly low 
complexity of the Lorenz system is comparable to that of limit cycles, 
e.g. Van der Pol oscillator.
The explanation for this comes about when we compute its complexity for shorter
time series. For example at 2000 points  the complexity is about $2.7$ times
higher than the complexity for a 10000 length series as reported in
Table~\ref{table2}. This phenomenon does not occur for the discrete system like
Henon attractor. Due to the fact that the Lorenz attractor contains a dense set
of unstable periodic orbits, long time evolution  affects the computation of
the complexity and reveals some resemblance  with periodic systems. Effects of
this magnitude did not appear in the  other time series analysed. In the 
deterministic intervals of the MG model, the complexity has a value comparable to 
that of the chaotic Henon map.  
   
The Lempel-Ziv complexity, as any other test or statistic, should always  
be used in conjunction with other diagnostic tools. In all simulations 
performed so far we have never found a stochastic process with complexity 
less than 0.8. However there are chaotic systems with complexity beyond this 
value, for example the family of maps $x_{n+1} = p x_n\,{\mbox{mod}}\,1$, 
$p$ a prime 
number. Another result is that known chaotic systems described by differential  
equations do not have high complexity. To conclude that a system is 
stochastic we employ recurrence plots and the cluster weighted modelling 
approach. The complexity is then used as a confirmation and, more importantly, 
to associate a level of stochasticity. This is done in the same way as using 
Lyapunov exponents to quantify chaos once determinism has been found.  

The recurrence plot is a visual method which helps in the identification  of
similarities and differences amongst diverse modes of  evolution. Several tests
with differential equations and maps,  represented here by the Lorenz and Henon
systems, show that it is  unlikely that low dimensional chaos can describe  
the kind of evolution found in the markets and in the MG model. In particular,
recurrences are clearly identified in the  Lorenz system and nothing  of the
kind will ever appear in stochastic models. In this sense stochastic processes 
are better suited to describe the financial index and the MG model. However, in 
the deterministic regime one can identify close return patterns similar to those 
of low dimensional chaotic systems. 

Cluster weighted analysis reveals another aspect of the minority game
behaviour. Its use in the modelling of general deterministic  systems produces
clusters whose variances are always smaller than the  variance of the data. In
contrast to this, when applied to a  stochastic system the variances of the
clusters are comparable  to the variances of the data. Such distinction is
preserved when we vary the number of clusters.

\section{Conclusions and Further Work}\label{sec:6}

The issue of nonstationarity is a subtle one. We limited ourselves  in this
study to unit root stationarity, but more sophisticated methods need to be used
for the several brands of MG models and financial indices. Using the complexity
parameter we found that long time evolution reveals some intrinsic features of
chaotic attractors described by differential equations. Also, this parameter
associated a higher complexity to the SP500 index as compared to the MG model
and a possible explanation for this was given above. The recurrence plots
confirm that, in general, the MG model cannot be described by low dimensional 
chaotic systems. The
nonlinear character of the model and the index  are clearly indicated by the
BDS test and surrogate data analysis. 

The MG model parameters employed in our simulations were chosen in the 
information efficient region. In the inefficient case our simulations show 
that the system is a non random process with complexity close to zero. The 
recurrence plot in this case provides a strong evidence of recurring 
orbits with several traces of horizontal segments. 
Thus, randomness is directly related to efficiency.  

An important finding in our analysis of the MG game is that low dimensional 
chaotic regimes are possible during the evolution. This occurs for intervals of 
about 1000 iterations, while shorter intervals of deterministic evolution are difficult to 
classify. A complete study of the probabilistic structure of the deterministic 
phases, and their statistical significance, requires a more extensive
investigation  \cite{ferreira:02-02}. For the full motion, comprised of
transitions between  deterministic modes induced by random perturbations, the
time series analysis has  indicated the operation of a nonlinear stochastic
process which is similar, but  not identical, to the SP500 index. In real
markets it is possible to find  periods of deterministic behaviour in exchange
rates for certain instruments and  specific periods of time \cite{degrawe:93}.
In the SP500 such occurrence is  unlikely but an extensive search in the data
base for several lengths and starting  points is feasible. Even if
deterministic modes can be found in historical  stock prices, indices or rates,
turning this into potential profit is a remote, but  not an entirely discarded,
possibility worthwhile pursuing.

Another interest in this investigation is that it provides a test amongst 
several models. In this sense one selects the models whose time series 
properties reproduce better real market evolution. In derivatives pricing 
a widely used test, and parameter estimation procedure, is to minimize the 
hedging error. This can be implemented only if a model is available, e.g. 
Black-Scholes. In a non-gaussian context there is no consensus on which 
model is appropriate. An extension of this work, which will be pursued in 
a future work, is to use time series methods in the several brands of agents 
based models and to compare them with market data. In addition to the 
stylized facts we should include nonlinearity, at least for  the SP500, 
and this is an additional reason to discard simple models based on the  
geometric Brownian motion. 

Additional investigations in this study is to model the trajectories of the 
Minority Game model having in view the pricing of derivatives instruments, and to 
use a more extensive set of statistics to examine other financial series 
in order to confront the similarities and differences of this model with 
real market data.

\ack{The authors acknowledge Damien Challet for useful e-mail on the implementation
of the grand canonical MG model. We acknowledge Funda\c{c}\~ao de Amparo \`a
Pesquisa do Estado de S\~ao Paulo of Brazil for fully  supporting this research
(F.F.F. and P.M.) and Coordenadoria de Aperfei\c{c}oamento de Pessoal de Ensino
Superior (B.S.M.)}

\begin{appendix}

\section{Description of the data sets}

The time series used as benchmarks were chosen to represent the kind of
behaviour we intend to identify in the evolution of the minority game model.
The Lorenz system and the Henon mapping  are prototypes of deterministic
behavior generated by differential equations and  differentiable mappings. ARMA
models and the NLMA are examples of linear and nonlinear stochastic processes. 
In the following we describe briefly the models used in the present study.

\subsection{Lorenz system:}
Nonlinear differential equations
\begin{align}\label{eq:lorenz}
\frac{dx}{dt} = \sigma (y-x);\;\;
\frac{dy}{dt} = -xz+rx-y;\;\;
\frac{dz}{dt}  = xy-\beta z
\end{align}
The parameter values are chosen as $\sigma =16$, $\beta =4$ and $r=40$. The
data series is obtained by solving Eqs.~(\ref{eq:lorenz}) numerically using
fourth order Runge-Kutta method.

\subsection{Henon map:}
Nonlinear differentiable mapping
\begin{align}
x_t = 1-ax_{t-1}^2+y_{t-1};\;\;
y_t = bx_{t-1}
\end{align}
where $a=1.4$ and $b=0.3$.

\subsection{Random (Noise): }

The data set is taken by drawing random points from a uniform distribution in
the unit interval $x_i\in \text{rand}(0,1)$.

\subsection{ARMA(2,1):}
Linear stochastic process
\begin{align}
x_t = 0.8x_{t-1}-0.5x_{t-2}+0.4\epsilon_{t-1}+\epsilon_t
\end{align}
where $\epsilon_t \sim i.i.d.$, a normal distribution with zero mean and
unit variance $N(0,1)$.

\subsection{NLMA(2):}
Nonlinear stochastic process
\begin{align}
y_t = \epsilon_t-0.3\epsilon_{t-1}+0.2\epsilon_{t-2}
+0.4\epsilon_{t-1}\epsilon_{t-2}-0.25\epsilon_{t-2}^2
\end{align}
where $\epsilon_t \sim i.i.d.$ $N(0,1)$.

\end{appendix}


\end{document}